\documentclass[a4paper,fleqn,usenatbib,useAMS]{mnras}

\usepackage{newtxtext,newtxmath}

\usepackage[T1]{fontenc}
\usepackage{ae,aecompl}

\usepackage{verbatim}
\usepackage{graphicx}	
\usepackage{amsmath}	
\usepackage{amssymb}	
\usepackage{stackengine}
\usepackage{xcolor}
\usepackage{soul}
\usepackage{times}
\usepackage{subfig}
\usepackage{breqn}
\usepackage{times}
\usepackage{epstopdf}
\usepackage{txfonts}
\usepackage{natbib}
\usepackage{array}

\title[Gravitational imaging of dark matter haloes]{Low-mass halo perturbations in strong gravitational lenses at redshift z$\sim$0.5 are consistent with CDM}

\author[E. Ritondale et al.]{
E. Ritondale$^{1}$\thanks{E-mail: elisa@mpa-graching.mpg.de},
S. Vegetti$^{1}$,
G. Despali$^{1}$,
M.~W. Auger$^{2}$,
L.~V.~E Koopmans$^{3}$,
\newauthor J.~P. McKean$^{3,4}$
 \\
$^{1}$Max Planck Institute for Astrophysics, Karl-Schwarzschild-Strasse 1, D-85740 Garching, Germany\\
$^{2}$Institute of Astronomy, Madingley Road, Cambridge CB30HA, UK\\
$^{3}$Kapteyn Astronomical Institute, University of Groningen, PO Box 800, NL-9700 AV Groningen, The Netherlands\\
$^{4}$Netherlands Institute for Radio Astronomy (ASTRON), PO Box 2, NL-7990 AA Dwingeloo, The Netherlands\\
}

\date{Accepted XXX. Received YYY; in original form ZZZ}

\pubyear{2015}

\begin{document}
\label{firstpage}
\pagerange{\pageref{firstpage}--\pageref{lastpage}}
\maketitle

\begin{abstract}
We use a sample of 17 strong gravitational lens systems from the BELLS GALLERY survey to quantify the amount of low-mass dark matter haloes within the lensing galaxies and along their lines of sight, and to constrain the properties of dark matter. Based on a detection criterion of 10$\sigma$, we report no significant detection in any of the lenses. Using the sensitivity function at the 10-$\sigma$ level, we have calculated the predicted number of detectable cold dark matter (CDM) line-of-sight haloes to be $\mu_{l} = 1.17\pm1.08$, in agreement with our null detection. Assuming a detection sensitivity that improved to the level implied by a 5-$\sigma$ threshold, the expected number of detectable line-of-sight haloes rises to $\mu_l = 9.0\pm3.0$. Whilst the current data find zero detections at this sensitivity level (which has a probability of P$^{{\rm5}\sigma}_{{\rm CDM}}(n_{\rm det}=0)$=0.0001 and would be in strong tension with the CDM framework), we find that such a low detection threshold leads to many spurious detections and non-detections and therefore the current lack of detections is unreliable and requires data with improved sensitivity. Combining this sample with a subsample of 11 SLACS lenses, we constrain the half-mode mass to be $\log$(M$_{\rm hm}) < 12.26$ at the 2-$\sigma$ level. The latter is consistent with resonantly produced sterile neutrino masses m$_{\rm s} < 0.8$ keV at any value of the lepton asymmetry at the 2-$\sigma$ level.
\end{abstract}

\begin{keywords}
gravitational lensing: strong -- galaxies: haloes -- galaxies: structure -- dark matter
\end{keywords}


%
\section{Introduction}
As much as 85  per cent of the matter content of the Universe is made of an unknown elusive component known as dark matter, and its nature represents one of the most long-standing and most studied problems in physics \citep[][]{Bosma1981,Rubin85,Frenk12}. In the standard Cold Dark Matter (CDM) cosmological model, this exotic matter component is described as made up of weakly interacting particles, such as axions and neutralinos, which have negligible thermal velocities at early times and are collisionless at scales smaller than $\sim$ 1 kpc \citep[e.g.][]{Baer15,Ringwald16}. Detailed observations of the Cosmic Microwave Background have shown that not long after the Big Bang the distribution of matter in the Universe was smooth and homogeneous except for small density perturbations \citep[][]{PlanckXII,PlanckXI,Schneider16}. Due to their negligible thermal velocity, CDM particles are confined within these fluctuations, which evolve under the influence of gravity and survive down to the smallest scales \citep[][]{Davis85, Yoshida00}. The distribution and evolution of these fluctuations determine the statistics of the dark matter distribution in the present Universe and constitute the backbone upon which baryonic matter builds up galaxies and galaxy clusters \citep[][]{Ostriker74,White78}.

The $\Lambda$CDM model is the cosmological framework that on large scales has provided the best agreement with observations to date \citep[e.g.][]{Springel2005, Planck16}. However, on smaller scales (less than a few kpcs), this agreement is less certain, and discrepancies between observations and predictions from high-resolution simulations arise \citep[e.g.][]{Kauffmann93,Moore94,BKolchin12}. In an attempt to alleviate these tensions, alternative dark matter models have been considered, for example, self-interacting and warm dark matter \citep[e.g.][]{Lovell14, Vogelsberger16, Robles17,Irsic17}. In particular, following the possible detection of a 3.5 keV line in the outskirts of the Perseus cluster, other nearby galaxy clusters \citep{Bulbul14}, the Andromeda galaxy \citep{Boyarsky14} and the Milky Way centre \citep{Boyarsky15}, sterile neutrinos with masses of a few keV have been proposed as one of the favoured alternative candidates \citep[however see][]{Anderson15,Jeltema16,Aharonian17}.

A critical difference between WDM and CDM models is the evolution of structure at galactic and sub-galactic scales. Thanks to their non-negligible velocities, WDM particles can freely stream out of small mass density perturbations in the early Universe, and are responsible for the suppression of the number density of low-mass dark matter haloes and subhaloes relative to CDM \citep[e.g.][]{Lovell12}. The mass scale at which this suppression happens depends on the momentum distribution of the dark matter particle and therefore its production mechanism. 
Quantifying the relative number of small-mass haloes is, therefore, an essential probe to the nature of dark matter. However, most of these haloes are expected to be very faint or even completely dark, and they can only be detected via their gravitational signature on the multiple images of gravitationally lensed quasars \citep{Mao98,DalalKochanek02,Nierenberg13,Gilman18} and lensed galaxies \citep[][]{Koopmans05, V09, V10, V12, V14a, Hezaveh16, Despali18,V18,Chatterjee18}, and the stellar distribution of globular-cluster streams in the Milky Way \citep[e.g.][]{Ngan14, Erkal16}.

In this paper, we apply the \emph{gravitational imaging} technique developed by \cite{Koopmans05} and \cite{V09} to a sample of 17 gravitational lenses from the BOSS Emission-Line Lens Survey (BELLS) for GALaxy-Ly$\alpha$ EmitteR sYstems \citep[BELLS GALLERY][]{Shu16a}. We combine the detections and non-detections of low-mass haloes to derive new statistical constraints on the dark matter mass function and compare our results with predictions from CDM and different sterile neutrino dark matter models. This paper is structured as follows: in Section~\ref{sec:data} we describe the data and in Section~\ref{sec:method} we present an overview of the adopted analysis scheme and summarise the gravitational imaging method. In Section~\ref{sec:stats}, we present the statistical approach used to infer the parameter of the subhalo and halo mass functions and the free-streaming properties of dark matter.
We give our results in Sections~\ref{sec:results} and \ref{sec:inferencedm}, and present our conclusions in Section~\ref{sec:conclusions}.

We assume the following cosmology throughout the paper, $H_0=71\; \mathrm{km\; s^{-1}\; Mpc^{-1}}$, $\rm{\Omega_m} = 0.27$ and $\rm{\Omega_{\Lambda}} = 0.73$.



\section{Data}
\label{sec:data}

The sample analysed in this paper consists of 17 galaxy-scale gravitational lens systems that were spectroscopically selected from the BELLS GALLERY of the Sloan Digital Sky Survey-III \citep{Shu16a}. The sample is both lens and source selected: 1.4 $\times$ 10$^6$ spectra were analysed to search for Lyman-$\alpha$ emission lines at a redshift higher than the foreground early-type-galaxy emission \citep{Shu16a}. Twenty-one lens candidates were then observed with the WFC3-UVIS camera and the F606W filter on board the {\it Hubble Space Telescope} ({\it HST}) between 2015 November and 2016 May (Cycle 23, program ID 14189, PI: A. Bolton). 

\begin{figure*}
\begin{center} 
\includegraphics[width= 17 cm, viewport=29 50 565 736, clip]{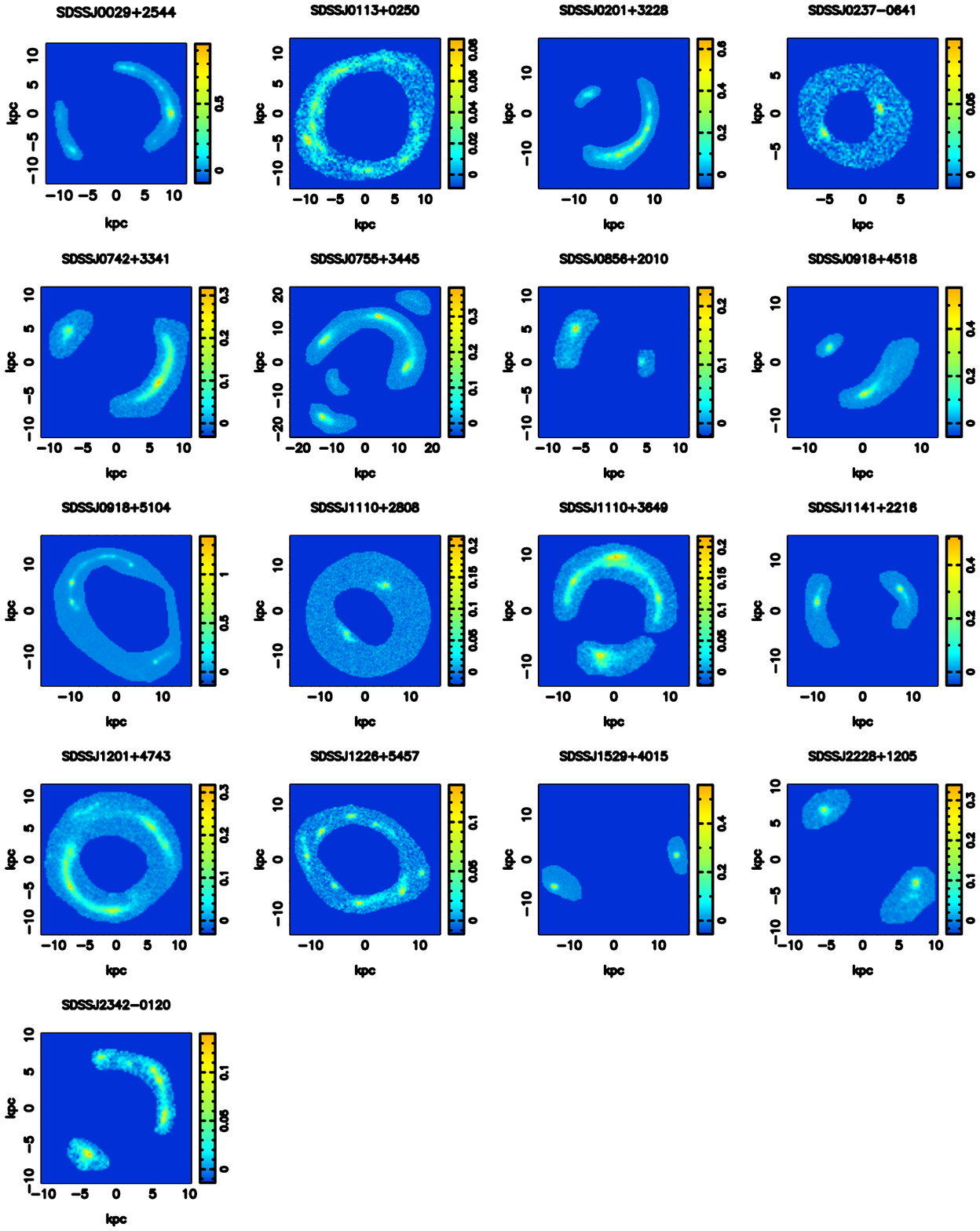}
\caption{The surface brightness distributions of the lensed images within a selected region on the sky plane used to reconstruct the background sources. The colour scale is in units of electron~s$^{-1}$ and the projected areas shown are at the redshift of the lens.}
\label{fig:sample}
\end{center}     
\end{figure*}

\begin{figure*}
\begin{center} 
\includegraphics[width= 17 cm, viewport=29 10 565 716, clip]{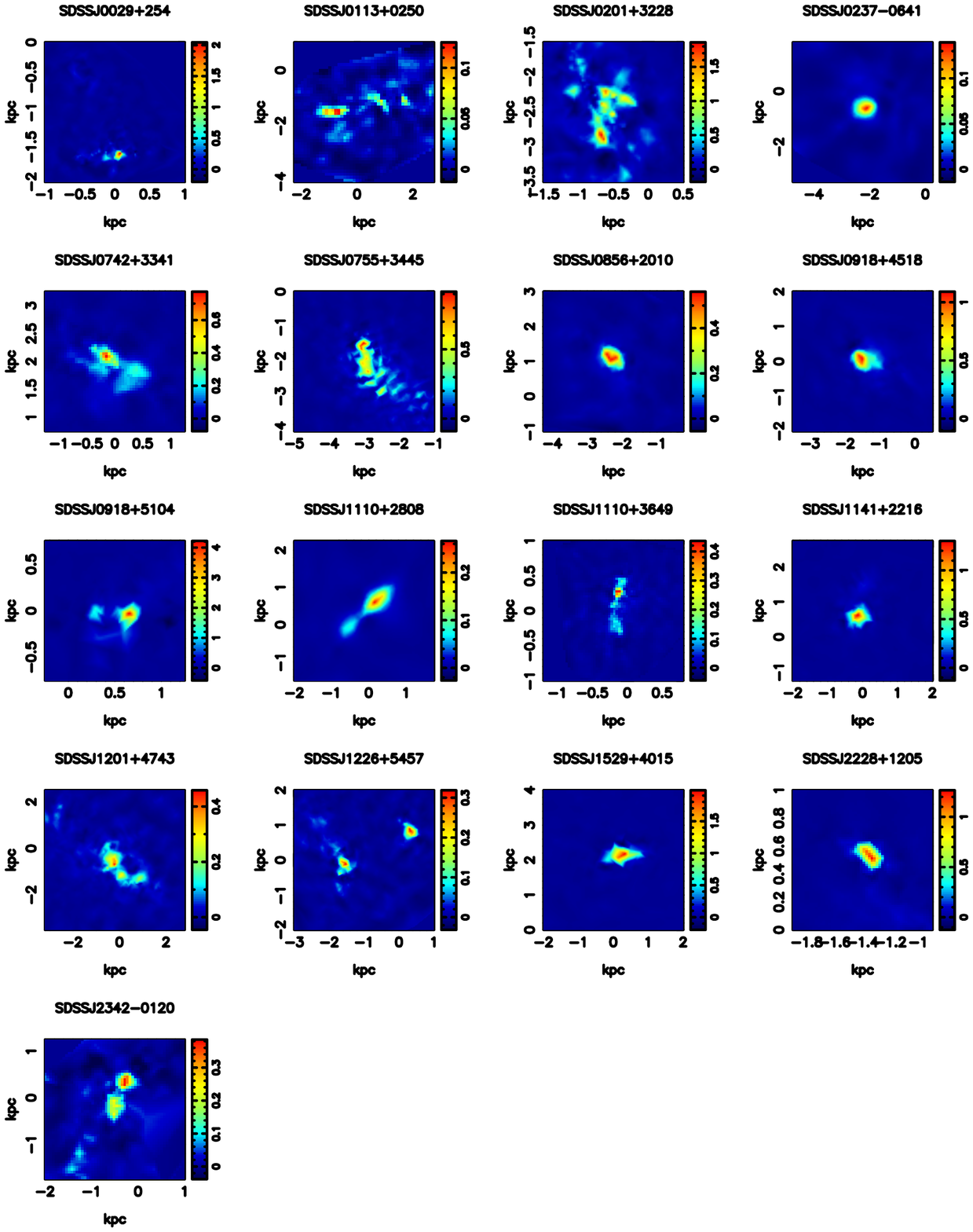}
\caption{The surface brightness of each background LAE, based on the pixelated source reconstructions from the gravitational lens modelling. The colour scale is in units of electron~s$^{-1}$ and the projected areas shown are at the redshift of the source.}
\label{fig:sourcepanel}
\end{center}     
\end{figure*}

Our final sample consists of 17 Lyman-$\alpha$ emitting galaxies at redshifts from 2.1 to 2.8 that are gravitationally lensed by massive early-type galaxies at a mean redshift of 0.5. For more details about the sample we refer to Table~\ref{tbl:list} and \cite{Shu16a}. Images of the lensed emission are shown in Fig.~\ref{fig:sample}. 
Gravitational lens models, under the assumption of a smooth elliptical power-law lensing potential, for all 17 systems can be found in \citet{Ritondale18} and more details can be found in Section~\ref{sec:smooth}.
Briefly,\citet{Ritondale18} revealed that these Lyman-$\alpha$ emitting sources are qualitatively very structured with extremely inhomogeneous surface brightness distributions. They often consist of multiple components that sometimes appear to be connected or merging, while in other cases, they appear as clearly separated components in the sky (see Fig.~\ref{fig:sourcepanel}). Their intrinsic sizes vary quite widely, from $0.2$~kpc to $1.8$~kpc in radius and they have a relatively low median integrated star formation rate of 1.4 M$_{\odot}$ yr$^{-1}$, on average \citep{Ritondale18}. 

\begin{table}
\begin{center}
\caption{Details of the gravitational lens systems analysed in this paper. For each system we list here the SDSS name, the lens and source redshifts, the rest-frame wavelength of the UV emission observed through the F606W filter, the observation exposure time and the Einstein radius from \citet{Ritondale18}. For lens systems with multiple lenses we list the Einstein radius for each deflector.}
\begin{tabular}{cccccc}
\hline
Name (SDSS) & $\rm{z_{\rm lens}}$ & $\rm{z_{\rm src}}$ & $\lambda_{\rm rest}$ & Exp. Time & $\rm{R_{\rm ein}}$\\
& & & [\AA] & [$s$]& [arcsec]\\
 \hline
$\rm{J0029+2544}$		&0.587	&2.450	&1706	&2504 &1.295\\
$\rm{J0113+0250}$		&0.623	&2.609	&1631	&2484 &1.226\\
	&&&&&0.065\\
	&&&&&0.172\\
$\rm{J0201+3228}$		&0.396	&2.821	&1540	&2520 &1.650\\
$\rm{J0237-0641}$		&0.486	&2.249	&1812	&2488 &0.687\\
$\rm{J0742+3341}$		&0.494	&2.363	&1751	&2520 &1.197\\
$\rm{J0755+3445}$		&0.722	&2.634	&1620	&2520 &1.926\\
$\rm{J0856+2010}$		&0.507	&2.233	&1821	&2496 &0.960\\
$\rm{J0918+4518}$		&0.581	&2.344	&1730	&2676 &0.444\\
&&&& &0.409\\
$\rm{J0918+5104}$		&0.581	&2.404	&1730	&2676 &1.600\\
$\rm{J1110+2808}$		&0.607	&2.399	&1732	&2504 &0.992 \\
$\rm{J1110+3649}$		&0.733	&2.502	&1682	&2540 &1.152\\
$\rm{J1141+2216}$		&0.586	&2.762	&1565	&2496 &1.281\\
$\rm{J1201+4743}$		&0.563	&2.126	&1883	&2624 &1.139\\
$\rm{J1226+5457}$		&0.498	&2.732	&1578	&2676 &1.351\\
$\rm{J1529+4015}$		&0.531	&2.792	&1553	&2580 &2.233\\
$\rm{J2228+1205}$		&0.530	&2.832	&1536	&2492 &1.291\\
$\rm{J2342-0120}$		&0.527	&2.265	&1803	&2484 &1.033\\
 \hline
\end{tabular}
\label{tbl:list} 
\end{center}
\end{table}

\section{Lens modelling}
\label{sec:method}

Each of the gravitational lens systems in the sample has been analysed with an updated version of the grid-based fully Bayesian modelling method developed by \cite{V09}. This technique simultaneously optimises for the surface brightness of the background source and the brightness and mass distribution of the foreground lens galaxy. Our lens modelling procedure is carried out in two subsequent steps. First, we find the best mass and light model for the main lensing galaxies under the assumption of a relatively smooth mass distribution, meaning that we do not allow for the presence of any subhalo or line-of-sight halo. Results of this analysis are presented and discussed by \citet{Ritondale18}. Then, we look for the gravitational signature of dark-matter subhaloes and line-of-sight haloes (collectively referred to as haloes for the rest of the paper) on the surface brightness distribution of the lensed images, and use their number and inferred masses to derive constraints on the dark matter properties. 
In this Section, we review the methodology, highlighting the main differences from the original version of \cite{V09}.

\subsection{Lens mass and light distribution model}

The mass distribution of the main lens(es) is parametrized as a cored elliptical power-law with normalised surface mass density $\kappa$ expressed as
\begin{equation}
\kappa\left (x,y \right ) = \frac{\kappa_0 \left ( 2 - \frac{\gamma}{2} \right ) q^{\gamma - 3/2}}{2 \left ( q^2 \left ( x^2 + r_c^2 \right ) + y^2\right )^{\left ( \gamma - 1 \right )/2}}\,,
\label{eq:massdensity}
\end{equation}
where $\kappa_0$ is the surface mass density normalization, $\gamma$ is the radial mass-density slope, $q$ is the axis ratio and $r_c$ the core-radius. The normalization is chosen to minimize the covariance between the parameters. The contribution of an external shear is parametrized by its strength $\Gamma$ and positional angle $\Gamma_\theta$.  

The foreground lens(es) light distribution is simultaneously optimised for and parametrized with a sum of elliptical S\'ersic profiles, each given by
\begin{equation}
S_i\left(x,y\right) = I_i \exp \left [ - a_i \left( \left (  \frac{\sqrt{q_{l,i}^2x^2+y^2}}{R_{e,i}} \right ) ^ {1/n_i}  - 1.0 \right)  \right ] = I_i~\Sigma_i\left(x,y\right) \,,
\label{eq:S\'ersic}
\end{equation}
where $I_i $ is the normalization, $a_i = 1.9992 \times n_i-0.3271$ \citep{Capaccioli89}, $q_{l,i}$ is the axis ratio, $R_{e,i}$ is the effective radius and $n_i$ is the S\'ersic index.
Given $j$ lenses and $i$ S\'ersic components, we refer to the free parameters of the mass and the light distributions collectively as $ \boldsymbol{\eta}_m =\{ k_{0,j}, \theta_j, q_j, x_j, y_j, r_{c,j}, \gamma_j, \Gamma_j, \Gamma_{\theta,j} \} $ and $\boldsymbol{\eta}_{l}=\{ R_{e,i}, n_i, x_l, y_l, q_{l,i},\theta_{l,i} \} $. They are all optimized, except for $r_c$, which we keep fixed at 0.1 mas. As the background source is also unknown, the other free parameters of the model are the source surface brightness distribution at each pixel on the source plane and regularisation level $\lambda_s$ that sets the  \emph{smoothness} of the source model \citep[see][for more details]{V09}. 

\subsection{Grid-based source model}
\label{sec:smooth}

The surface brightness distribution of each pixel in the lens plane $\mathbf{d}$ is given by the sum of the lensed emission $\bmath{d_s}$ of the background source $\bmath{s}$ and the foreground lens brightness distribution $\mathbf{d}_l $. The positions of the pixels on the lens- and source-planes are related to each other by the lensing potential $\psi_s \left(\mathbfit{x}, \boldsymbol{\eta}_m \right)$ via the lens equation, where we consider one lens and one source plane respectively. The effect of the line-of-sight haloes is projected onto the main lens plane, using the mass-redshift relation by \cite{Despali18} (see also Section~\ref{sec:clumpy}), which takes into account the non-linear effects due to the multi-plane lens configuration. The conservation of surface brightness by gravitational lensing then leads to the following set of linear equations
\begin{multline}
\mathbf{B}\left[\mathbf{L}(\boldsymbol{\psi}, \boldsymbol{\eta}_m )~|~\left(\bmath{\Sigma}_0~...~\bmath{\Sigma}_n\right)~|~\bmath{1}\right]\left(\begin{array}{c}\bmath s\\ I_0\\ .\\.\\.\\I_n\\b\end{array} \right)+\bmath{n} = \bmath{d_s}+\bmath{d_l} = \mathbf{d}\,,
\label{eq:linear_system}
\end{multline}
where $\mathbf{B} $ is the blurring operator, which represents the effect of the point spread function, $\mathbf{L} $ is the lensing operator, $\bmath{\Sigma_i}$ is the surface brightness at each pixel of the $i$th S\'ersic component and $\bmath{n}$ is the noise vector. The values of $I_0,...,I_{n}$ are the unknown S\'ersic component normalizations and $b$ is a constant accounting for the sky background. Under the assumption of Gaussian noise, the maximum a posteriori (MAP) parameters $ \boldsymbol{\eta}_m $, $\lambda_s$ and $ \boldsymbol{\eta}_l$ can be inferred by maximizing the penalty function
\begin{equation}
P\left(\mathbf{r} , \boldsymbol{\eta}_m, \boldsymbol{\eta}_l, \lambda_s~|~\mathbf{d}, \mathbf{H}_s\right) \propto \|\mathbf{M} \mathbf{r}  - \mathbf{d} \|_2^2 + \lambda_s^{2} \|\mathbf{H}_s \mathbf{r} \|_2^2\,,
\label{eq:penalty} 
\end{equation}
where we have introduced the vector $  \mathbf{r}^\intercal = \left( \bmath{s}, \\ I_0\\ .\\.\\.\\I_n\\, b \right)$ and rephrased equation~(\ref{eq:linear_system}) as 
\begin{equation}
\mathbf{M} \mathbf{r}  + \mathbf{n} = \mathbf{d} \,,
\label{eq:operator} 
 \end{equation}
with $\mathbf{M}$ the product of the blurring operator with the lensing operator and foreground surface brightness matrix. The first term of the penalty function represents the $\chi^2$ difference between the data and the model, while the second term includes a priori information on the smoothness of the source surface brightness distribution encoded by the level and form of the regularization $\lambda_s$ and $\mathbf{H}_s$. The latter is set to zero in the entries multiplying the S\'ersic component normalizations. For each choice of the non-linear parameters $\boldsymbol{\eta}_m$,  $\boldsymbol{\eta}_l$ and $\lambda_s$, the corresponding value for $\mathbf{r} $ is obtained by solving the linear system
\begin{equation}
\left( \mathbf{M}^T  \mathbf{C}_{d}^{-1} \mathbf{M}  + \lambda_s^{2}~\mathbf{H}_s^T \mathbf{H}_s\right)\mathbf{r}  = \mathbf{M}^T  \mathbf{C}_{d}^{-1} \mathbf{d} \,,
\label{eq:source} 
\end{equation}
where $\mathbf{C}_{d}$ is the data covariance matrix and is assumed to be diagonal, that is, we ignore any correlation among data pixels. In fact the only non-zero correlation term is due to drizzling, which is as low as 0.2 between adjacent pixels and effectively zero up to the second neighbouring pixel \citep{Bayer18}. 

We refer the reader to \citet{Ritondale18} for a more detailed description of the smooth modelling results for the sample of lenses studied in this paper.

\subsection{Grid-based Potential Corrections}
\label{sec:potcorr}

At this stage of the lens modelling, we gravitationally image low-mass substructures by describing them as linear localized pixellated corrections $\delta \psi(\mathbf{x})$ to the main lensing potential. In practice, at each position $\mathbf{x}$ on the main lens plane, we redefine the lensing potential as $\psi(\mathbf{x}, \boldsymbol{\eta}_m) = \psi_s(\mathbf{x},\boldsymbol{\eta}_m) + \delta \psi(\mathbf{x})$. Where $\psi_s(\mathbf{x},\boldsymbol{\eta}_m)$ is the parametric smooth potential introduced in the previous section. Following the formalism developed by \cite {Koopmans05} and \cite{V09}, we then introduce a new linear system relating the image and the source planes, which at each iteration $n$ reads as 
\begin{equation}
\mathbf{M} \left(\boldsymbol{\eta}_m, \boldsymbol{\eta}_l, \boldsymbol{\psi}_{n-1},\bmath{s}_{n-1}\right) \mathbf{r}_{n} + \mathbf{n} = \mathbf{d} \,,
\label{eq:linear_system_potcorr}
\end{equation}
with
\begin{equation}
\mathbf{M}  = \mathbf{B} \left[\mathbf{L} \left(\boldsymbol{\eta}_m,\boldsymbol{\psi}_{n-1}\right)~|~-\mathbf{D}_s(s_{n-1})\mathbf{D}_{\boldsymbol{\psi}}~|~\bmath{\Sigma_i}~|~\bmath{1}\right]\,
\end{equation}
and
\begin{equation}
 \mathbf{r}_{n}^\intercal = \left( \bmath{s}, \\ I_0\\ .\\.\\.\\I_n\\, b, \delta \boldsymbol{\psi}_n \right).
\end{equation}
Here, $\mathbf{D}_s(s_{n-1})$ is a sparse matrix whose entries depend on the surface brightness gradient of the best source at the $n-1$ iteration and $\mathbf{D}_{\boldsymbol{\psi}}$ is a matrix that
determines the gradient of $\delta \boldsymbol{\psi}_n$ \citep[see][for more details]{Koopmans05}. Introducing $ \mathbf{H}_{\delta \boldsymbol{\psi}}$ and $\lambda_{\delta \boldsymbol{\psi}}$ as the form and level of regularization for the potential corrections $\delta \boldsymbol{\psi}_n$, we can now write a new penalty function as
\begin{multline}
P\left(\mathbf{r}_n~|~\mathbf{d} , \boldsymbol{\eta}_m, \boldsymbol{\eta}_l, \lambda_s,\lambda_{\delta \boldsymbol{\psi}}, \bmath{s}_{n-1}, \mathbf{H}_s, \mathbf{H}_{\delta \boldsymbol{\psi}}\right) \propto\\
\|\mathbf{M} \mathbf{r}_n - \mathbf{d} \|_2^2 + \lambda_s^{2} \|\mathbf{H}_s\bmath{s}_n\|_2^2 + \lambda_{\delta \boldsymbol{\psi}}^{2} \|\mathbf{H}_{\delta \boldsymbol{\psi}}\delta \boldsymbol{\psi}_n\|_2^2 \,.
\label{eq:penalty_potcorr} 
\end{multline}
We can further define $\mathbf{R}$ as the diagonal block matrix that contains the regularization level parameters $\lambda_s$ and $\lambda_{\delta \boldsymbol{\psi}}$, and combines the source and potential regularization operators $\mathbf{H}_s$ and $\mathbf{H}_{\delta \boldsymbol{\psi}}$. Maximising the penalty function with respect to $\mathbf{r}$ leads to the following set of equations for $\mathbf{r}_n$
\begin{equation}
\left(\mathbf{M}^T\mathbf{C}_d^{-1}\mathbf{M} +\mathbf{R} \right)\mathbf{r}_n=\mathbf{M}^T\mathbf{C}_d^{-1}\mathbf{d}. 
\label{equ:linear_potcorr}
\end{equation}
 The solution of this linear system can be found using an iterative technique; in particular, we solve equation (\ref{equ:linear_potcorr}) and then add the correction $\delta \boldsymbol{\psi}_n$ to the best potential of the previous iteration $\boldsymbol{\psi}_{n-1}$. While iterating this procedure, both the source and the potential should converge to the maximum of the penalty function, given by equation~(\ref{eq:penalty_potcorr}), following a Gauss-Newton scheme. At every step of this procedure, the matrix $\mathbf{M}$ has to be recalculated for the new updated potential $\boldsymbol{\psi}_n$ and source $\bmath{s}_n$. While the potential grid points are kept spatially fixed in the image plane, a Delaunay tessellation grid for the source is rebuilt at every iteration to ensure that the number of degrees of freedom is kept constant during the entire optimization process. Pixellated convergence corrections can be derived from the corresponding potential corrections by applying the Laplace operator. At this stage of the analysis, the non-linear parameters $\boldsymbol{\eta}_m$, $\boldsymbol{\eta}_l$ and $\lambda_s$ are kept fixed at the MAP parameters inferred in Section~\ref{sec:smooth}, which means that $\delta \boldsymbol{\psi}$ includes corrections due to substructure as well as any departures from the macro-model assumptions. This is an important aspect of this technique that allows us to distinguish between genuine and spurious substructure detections (Ritondale et al., in prep). However, a systematic quantification of the interplay and degeneracy between complex mass distribution and the properties of substructure and of the background source has not been studied yet and we plan to address this issue in a follow-up paper.

\subsection{Small mass haloes as analytical mass components}
\label{sec:clumpy}

The main advantage of describing substructures as pixellated potential corrections is that it does not require any prior assumption on their number and mass density profile.
However, the non-linear parameters describing the main lensing potential are fixed at the best smooth values and the number of degrees of freedom defined by the potential correction grid can be relatively large. Therefore, the \textit{gravitational imaging} alone does not allow the degeneracy between the properties of the main lens and those of the substructure to be straightforwardly quantified nor to be used to statistically compare models \citep{V09}. 

To this end, we follow our pixellated analysis with an analytical description of the mass density profile of the low-mass haloes. At this stage of the analysis, we assume that all of the haloes are at the same redshift of the lens, that is subhaloes. Using the mass-redshift relation from \cite{Despali18}, we then derive the mass that these haloes should have had in order to generate the most similar lensing effect if they were located at a different redshift, that is, along the line of sight. In this procedure, we take into account the non-linear effects due to the multi-plane lens configuration. At this step, the haloes are parametrized by a spherically symmetric NFW profile \citep{Navarro96} with the concentration-mass relation of \cite{Duffy08}. While this mass density profile is a good description for the line-of-sight haloes, it is only an approximation for the subhaloes that have been accreted by the halo of the lens galaxy, and have therefore experienced events of tidal disruption. At a fixed virial mass, this results in a higher concentration that is mildly dependent on the subhalo distance from the host centre. However, \cite{Despali18} have shown that assuming a constant mass-concentration relation from \cite{Duffy08} plays only a secondary effect and leads to an error on the inferred mass of 5 per cent for subhaloes with masses of $10^{5-6}M_\odot$ and 20 per cent for masses of $10^9M_\odot$. These errors are significantly smaller than the de-projection errors on the total mass of pseudo-Jaffe profiles \citep[][]{Minor16, Despali18} and it leads to an error on the expected number of substructures of the order of 10 per cent. 

At this stage of the analysis, the free parameters of the model are: the non-linear parameters describing the main lens mass and light distribution, the source surface brightness distribution in each pixel and its regularization, the NFW virial mass and the projected position of each halo.

\subsection{Bayesian evidence and model comparison}
\label{sec:evidence}

In order to determine the statistical significance of a substructure detection, we compare the marginalised Bayesian evidence of the smooth and perturbed analytical models to determine which of the two is preferred by the data. For each system, the Bayesian evidence is computed using {\sc MultiNest} \citep{Multinest13} as the following integral
\begin{multline}
\mathcal{E} = P\left(\mathbf{d} |\mathbf{M} , \mathbf{H}_s\right) = \\
\int P\left(\mathbf{d} | \lambda_s, \boldsymbol{\eta}_m, \boldsymbol{\eta}_l, m, \mathbf{x}, \mathbf{M}, \mathbf{H}_s\right) P\left(\lambda_s, \boldsymbol{\eta}_m, \boldsymbol{\eta}_l, m, \mathbf{x}\right) d\lambda_s d \boldsymbol{\eta}_m d \boldsymbol{\eta}_l d m d \mathbf{x}\,, 
\label{eq:int_evidence}
\end{multline}
where, $P\left(\lambda_s, \boldsymbol{\eta}_m, \boldsymbol{\eta}_l, m, \mathbf{x}\right)$ is the normalized prior probability density distribution on the model parameters, and is chosen as follows: for the non-linear parameters $\boldsymbol{\eta}_m$, $\boldsymbol{\eta}_l$ and $\lambda_s$, we choose uniform priors within an interval centred on the MAP value derived in Section \ref{sec:method} and as large as 10, 20 or 40 per cent of this value, with priors always consistent between the smooth and the perturbed model for each lens \footnote{Building the prior volume based on the data is not consistent with a Bayesian approach, but we checked that this does not impact the resulting parameter values.}. For the source regularization level $\lambda_s$,  the prior is uniform in logarithmic space. Both the prior on the model parameters and the likelihood are properly normalized to have integrals of unity. At this stage, substructures are described analytically (as discussed in Section \ref{sec:clumpy}). Their masses $m$ have a uniform prior in logarithmic space, while their positions $\mathbf{x}$ are equally probable at any location  on the plane of the lensed images.

\subsubsection{Detection criteria}
\label{sec:criteria}

As demonstrated by \citet{McKean07}, \citet{Gilman17a} and \citet{Hsueh16,Hsueh17,Hsueh18}, assuming that all departures from a smooth power-law elliptical potential are due to the presence of dark sub-haloes and line-of-sight haloes can lead to the false detection of haloes with a high statistical significance. Indeed, a complex lensing potential, as for example in the form of edge-on discs, can affect the lensed observables in a way which is degenerate with a large number of low-mass haloes. In this respect, the pixellated gravitational imaging technique, described in Section \ref{sec:potcorr}, represents a clear advantage as it allows for the identification and quantification of all departures from a simple power-law macro model, independently of their origin. To obtain a reliable set of detections and non-detections, it is therefore important to combine the results of the two analyses that have been carried out in a completely independent way. Following \cite{V14a}, we define a detection as robust if:

\begin{enumerate}

\item a positive and localized convergence correction is identified, it improves the fitting quality of the data and does not depend on the source regularization forms and levels;\\

\item the analysis using parametric models for the haloes in Sections \ref{sec:clumpy} and \ref{sec:evidence} leads to the detection of a substructure with the same mass and at the same location as the peak of the convergence corrections identified at the previous step;\\

\item the model that includes the presence of a substructure is preferred over the smooth model by a difference in the Bayesian evidence of $\Delta \log \mathcal{E} = \log \mathcal{E}_{\rm pert} - \log \mathcal{E}_{\rm sm} \geq 50$, corresponding roughly to a 10-$\sigma$ detection at its inferred position. 

\end{enumerate}

\subsubsection{Detection threshold}
\label{sec:tests}
As described in the previous section, we choose to set our detection threshold at a Bayes factor of $\Delta\log \mathcal{E} \geq 50$. Under the assumption of statistical Gaussian errors, this corresponds to a 10$\sigma$-threshold. Using the reconstructed sources from the 10 systems with the highest number of data pixels in the lens plane, we have also tested the reliability of lower-significance cuts. In particular, from each of the 10 reconstructed sources we have created four mock lensed data sets with the same level of signal-to-noise ratio and resolution as the original data: one smooth model and three including a subhalo detectable at the 4 ($\Delta\log \mathcal{E} \geq 8$), 5 ($\Delta\log \mathcal{E} \geq 12.5$) and 6-$\sigma$ ($\Delta\log \mathcal{E} \geq 18$) level. We have then analysed these data in the same way as the real dataset and overall found a high percentage of false positive (almost 60 per cent) and false negatives (almost 40 per cent). In 30 per cent of the cases we have either recovered the correct lack of subhaloes or correctly detected the presence of an existing one. The percentage of false positives and false negatives drops from 100 per cent (40 and 60 per cent respectively) at the 4-$\sigma$ level, to 90 per cent (20 and 70 per cent respectively) at the 5-$\sigma$ one, and finally to 60 per cent (30 per cent for both) at the 6-$\sigma$ case. We therefore conclude that these lower-significance cuts are statistically unreliable. We believe this to be related to the fact that the conversion between the Bayes factor and a simple confidence level is only valid under the approximation of Gaussian statistical errors and does not include the effects of systematics (especially in relation to the source structure). In the rest of the paper, we therefore quote as robust our results based on the more conservative 10-$\sigma$ cut, for which we recover correct results in 80 per cent of the cases (10 per cent of false positives and 10 per cent of false negatives). 

\section{Inference on dark matter}
\label{sec:stats}
In this section, we describe how the detection and non-detection of subhaloes and line-of-sight haloes are statistically combined to constrain the free-streaming properties of dark matter. 

\subsection{Mass and position definition}
\label{sec:mass_def}

In the following, we denote with $m^{\rm o}$ the observed NFW virial mass, that is the mass that one would infer from the lens modelling of the data under the substructure assumption. This mass is allowed to vary between the lowest detectable mass $M_{{\rm low}}^{{\rm NFW}}(\mathbf{x}^{\rm o})$ at each considered projected position $\mathbf{x}^{\rm o}$ (see Section \ref{sec:sensitivity} for a definition) and the maximum NFW virial mass $M_{{\rm max}}^{{\rm NFW}} = 10^{11}$ M$_{\odot}$. 
The true NFW virial mass of a substructure at the redshift of the lens or a line-of-sight halo at an arbitrary redshift $z$ is referred to as $m$, and it is allowed to assume any value between $M^{\rm NFW}_{\rm min}= 1.0\times10^{5}$ M$_\odot$ and  $M^{\rm NFW}_{\rm max}$. The observed and true mass are statistically related to each other via equation~(\ref{eq:loslenseff}). The observed and true projected position of haloes are respectively indicated with $\mathbf{x}^{\rm o}$ and $\mathbf{x}$. The former is defined as the projected position on the lens plane where the lensed images are affected by the presence of the halo. For subhaloes, $\mathbf{x}^{\rm o}$ and $\mathbf{x}$ are related to each other by a relatively small measurement error \citep{Despali18}. For line-of-sight haloes, the recursive nature of the lens equation needs to be taken into account.

\subsection{Dark matter mass function}
\label{sec:massfunction}

Following \cite{Schneider12} and \cite{Lovell14}, we parametrise the subhalo and halo mass function as
\begin{equation}
n \left(m|M_{\rm{hm}},\beta \right) = n^{\rm{CDM}}\left(m\right) \left(1+ \frac{M_{\rm{hm}}}{m} \right)^{\beta}\,,
\label{eq:dndmlos}
\end{equation}
where $n^{\rm{CDM}}\left(m\right)$ is the number density of objects with mass $m$ in the CDM framework and the second factor expresses the suppression in the number of haloes due to the free streaming of the dark matter particles. In particular, $M_{\rm{hm}}$ is the mass scale at which the WDM mass power-spectrum is suppressed by one half with respect to the CDM one. For the subhalo CDM mass function we assume a power-law mass function given by
\begin{equation}
n_{\rm{sub}}^{\rm{CDM}}\left(m\right) \propto m^{-\alpha}.
\label{eq:dndm_sub}
\end{equation}
Instead, for the CDM mass function of line-of-sight haloes, we adopt the expression by \cite{Sheth99b}, with the best-fit parameters optimized for the Planck cosmology from \cite{Despali16} \footnote{These cosmology parameters are slightly different ($<$10 per cent) from the ones stated at the beginning of the paper. However, this difference does not impact the formation of structures as small as the ones we are concerned with. Therefore, this difference is not important for our purposes \citep{Despali16}.}.

\subsection{Likelihood}
\label{sec:likelihood}

In the following, we refer to $m^{\rm ob}_i$ and $\mathbf{x}^{\rm ob}_i$ as the bins of observed mass $m^{\rm o}$ and projected position $\mathbf{x}^{\rm o}$ that correspond to a detected halo. As in \cite{V18}, we have chosen the widths of these mass and position bins to be small enough so that the maximum number of detections per bin is one. We have also assumed a Poisson distribution for the number of haloes. Under this assumption, we can express the likelihood of detecting $n$ objects (substructures plus line-of-sight haloes) with observed NFW masses $\left\{m^{\rm ob}_1, ...., m^{\rm ob}_n\right\}$ at the projected positions  $\left\{{\bf x}^{\rm ob}_1, ...., {\bf x}^{\rm ob}_n\right\}$, and no detection in all other mass and position ranges as follows \citep[see][for a complete derivation]{V18}
\begin{multline}
\log P\left(\left\{m^{\rm ob}_1, ....,m^{\rm ob}_n\right\},\left\{{\bf x}^{\rm ob}_1, ....,{\bf x}^{\rm ob}_n\right\}|\theta\right) =  \\
-\int{\left[\mu_s(m^{\rm o},\mathbf{x}^{\rm o})+\mu_l(m^{\rm o},\mathbf{x}^{\rm o})\right]dm^{\rm o}d\mathbf{x}^{\rm o}}+\\
\sum_i^n \log\left[\mu_s\left(m_i^{\rm ob},\mathbf{x}^{\rm ob}_i\right)dm^{\rm o}d\mathbf{x}^{\rm o}+\mu_l\left(m_i^{\rm ob},\mathbf{x}^{\rm ob}_i\right)dm^{\rm o}d\mathbf{x}^{\rm o}\right]\,,
\label{eq:likelihood}
\end{multline}
where $\boldsymbol{\theta}$ is a vector containing the set of parameters describing the subhalo and halo mass functions (see Section~\ref{sec:prior} for an explicit definition). The above integrals are computed between the lowest detectable mass $M_{{\rm low}}^{{\rm NFW}}(\mathbf{x}^{\rm o})$ and $M_{{\rm max}}^{{\rm NFW}}$, while for the positions we consider all pixels on the lens plane used to reconstruct the background source (see Fig.~\ref{fig:sample}).
Here, $\mu_s \left(m^{\rm o},\mathbf{x}^{\rm o}\right)dm^{\rm o}d\mathbf{x}^{\rm o}$  and $\mu_l\left(m^{\rm o},\mathbf{x}^{\rm o}\right)dm^{\rm o}d\mathbf{x}^{\rm o}$ are the mean expected number of substructures and line-of-sight haloes, respectively, in the mass range $m^{\rm o},m^{\rm o}+dm^{\rm o}$, and projected position range ${\bf x}^{\rm o},{\bf x}^{\rm o}+ d{\bf x}^{\rm o}$. These are derived in Section \ref{sec:expectval}.

\subsection{Sensitivity function}
\label{sec:sensitivity}

In order to derive constraints on the substructure and halo mass function, it is necessary to calculate the sensitivity function for each lens system in the sample. The latter is defined as the lowest detectable NFW mass at the redshift of the main lens $M_{{\rm low}}^{{\rm NFW}}(\mathbf{x}^{\rm o})$ for each position in the lens plane of each system in the sample. It is computed as the smallest observed mass for which a clumpy model is preferred over the smooth one by a factor of the marginalized Bayesian evidence (see Section~\ref{sec:evidence}) corresponding to a 10-$\sigma$ detection cut.

As demonstrated by \cite{Koopmans05} and \cite{Rau14}, the sensitivity function strongly depends on the complexity of the surface brightness distribution of the background source. In fact, the strength of a surface brightness anomaly ($\delta$I) due to a potential perturbation $\delta \boldsymbol{\psi}$ is  $\delta I = - \nabla \bmath{s}  \cdot \nabla \delta \boldsymbol{\psi}$, that is, the inner product of the gradient of the source brightness distribution (where $\nabla \bmath{s}$ is evaluated in the source plane) dotted with the gradient of the potential perturbation due to substructure (where $\delta \boldsymbol{\psi}$ is evaluated in the image plane). Hence, mass (sub)structure can be detected more easily for sources that are highly structured (i.e. large values of $\left | \nabla \bmath{s} \right |$) or, conversely, more structured sources allow for a lower mass detection threshold for a fixed signal-to-noise ratio.

The sample analyzed in this paper consists of 17 lensed Lyman-$\alpha$ emitters, that are known to be very structured galaxies and characterised by a high dynamical range in their brightness distribution \citep{Ritondale18}. Therefore, these data could in principle be characterized by a relatively high sensitivity (i.e. low mass detection threshold). We present the actual distribution of pixels as a function of lowest-detectable mass in comparison with the SLACS lenses in Section~\ref{sec:mu}.

\subsection{Expectation values}
\label{sec:expectval}

Given the sensitivity functions, it is now possible to compute the expected total number of subhaloes and line-of-sight haloes as 
\begin{multline}
\mu(m^{\rm o},\mathbf{x}^{\rm o}) = \mu_{0} \times\int P(I=1|m^{\rm o},\mathbf{x}^{\rm o})P(m^{\rm o}|m,z)\\P(m,z |\boldsymbol{\theta})P(\mathbf{x}^{\rm o}|\mathbf{x},z)P(\mathbf{x})P(z)~dmdz d\mathbf{x} \,.
\label{eq:mu}
\end{multline}
The mass integrals are evaluated between $M^{\rm NFW}_{\rm min}$ and $M^{\rm NFW}_{\rm max}$. $I$ is a vector that is equal to one for detectable objects and zero otherwise, so that $P(I=1|m^{\rm o},\mathbf{x}^{\rm o})$ encodes the sensitivity function and is given by
\begin{equation}
P(I=1|m^{\rm o},\mathbf{x}^{\rm o}) = \begin{cases} 1 & \mbox{if} ~m^{\rm o} \geq M_{\rm low}(\mathbf{x}^{\rm o})\\ 0 & \mbox{otherwise} \end{cases}\,.
\end{equation}
$P(m,z |\theta)dmdz$ is the probability of finding one halo in the mass range $m,m+dm$ and in the redshift range $z,z+dz$. It is related to the mass functions by
\begin{equation}
P(m,z|\boldsymbol{\theta})dmdz = n(m,z|\boldsymbol{\theta})\frac{dV}{dz}dmdz\left[\int n(m^\prime,z^\prime|\boldsymbol{\theta})\frac{dV}{dz^\prime}dm^\prime dz^\prime\right]^{-1}\,,
\end{equation}
and, for subhaloes, it reduces to the following expression 
\begin{equation}
P(m|\boldsymbol{\theta}){\bf dm} = n(m|\boldsymbol{\theta}){\bf dm}\left[\int{ n(m^\prime|\boldsymbol{\theta})dm^\prime}\right]^{-1}
\end{equation}
with $n(m|\boldsymbol{\theta})$ given by equation~(\ref{eq:dndmlos}).

Introducing the projected mass of the main lens $M_{\rm lens}$ and a projected total mass fraction in substructure $f_{\rm sub}$ between $M^{\rm NFW}_{\rm min} $ and  $M^{\rm NFW}_{\rm max}$, we express $\mu_{0}$ as follows
\begin{equation}
\mu_{0} = f_{\rm sub} M_{\rm lens}\left[\int{m^\prime P(m^\prime|\boldsymbol{\theta})dm^\prime}\right]^{-1}\,.
\label{equ:mu_s}
\end{equation}
For line-of-sight haloes $\mu_{0}$ is expressed instead as 
\begin{equation}
\mu_{0} = \int n(m^\prime,z^\prime|\boldsymbol{\theta})\frac{dV(\mathbf{x}^\prime)}{dz^\prime}dm^\prime dz^\prime d\mathbf{x}^\prime\,.
\end{equation}
As the measurement error on the halo positions is relatively small \citep{Despali18}, we assume $P(\mathbf{x}^{\rm o}|\mathbf{x},z)=\delta(\mathbf{x}-g(\mathbf{x}^{\rm o},z))$, that is a delta function. For substructure, $g(\mathbf{x}^{\rm o},z)\equiv \mathbf{x}^{\rm o}$, whilst for line-of-sight haloes $g(\mathbf{x}^{\rm o},z)$ takes into account the effect of the recursive lens equation. Following the results by \citet{Xu15} and \citet{Despali18}, we assume a uniform probability for $P(\mathbf{x})$. The redshift of line-of-sight haloes have a uniform prior between the observer and the source, excluding the region within the main lens virial radius. The latter estimated to be $\sim$ 390 kpc and 0.0001 in redshift, assuming that the lens galaxies are typical early-types at $z \sim 0.5$ and have a mean halo mass of $~\rm{M=10^{13}~M_{\odot}}$. For subhaloes $P(z)=\delta\left(z-z_{\rm lens}\right)$.

As the detection threshold  $M_{{\rm low}}^{{\rm NFW}}(\mathbf{x}^{\rm o})$ and the measured mass $m^{\rm o}$ were derived under the substructure assumption, we account for the different lensing effect of line-of-sight haloes via the term
\begin{equation}
P(m^{\rm o}|m,z) = \frac{1}{\sqrt{2 \pi}m^{\rm o}\sigma(z)}\exp{\left[-\frac{\left(\log m^{\rm o}-f(m,z)\right)^2}{2\sigma^2(z)}\right]}\,.
\label{eq:loslenseff}
\end{equation}
Given a line-of-sight halo of NFW virial mass $m$ located at redshift \textit{z}, $f(m, z)$ returns the NFW virial mass at the same redshift of the main lens with the most similar gravitational lensing effect, for a \citet{Duffy08} concentration-mass relation. Here, we do not use the exact relation reported by \citet{Despali18}, but we derive a characteristic relation for each lens in our sample using mock lensed images of each of the lenses in the sample. The intrinsic scatter $\sigma(z)$ is also not the same as the one given by \citet{Despali18}, but it is a sum in quadrature of the error on the measured mass $m^{\rm o}$ and the uncertainty related to changes in the mass redshift-relation as a function of position on the image plane in a way that depends on the main lens deflection angle and the external shear. In the case of substructures, $f(m,z)$ reduces to $m$ and $\sigma(z)$ reduces to the mass measurement error.  

\subsection{Prior and posterior distributions}
\label{sec:prior}

The target parameters of the model, expressed by the vector  $\boldsymbol{\theta}$, are the subhalo and halo mass function slopes, respectively $\alpha$ and $\beta$, the projected total dark matter fraction in substructures $f_{\rm sub}$, and the half-mode mass $M_{\rm hm}$. These are drawn from the following prior probabililty density distributions:
(i) for $\alpha$ and $\beta$ we assume a Gaussian prior centred at 1.9 and $-$1.3 and with $\sigma$ of 0.2 and 0.1, respectively;
(ii) we draw the values of $f_{\rm sub}$ from a uniform prior density distribution in $1/\sqrt{f_{\rm sub}}$ between 0 and 0.2; (iii) for the half-mode mass we assume a logarithmic prior distribution between $10^6$ and $2\times 10^{12}$ M$_{\odot}$. Both priors are chosen in order to allow for an even exploration of the parameter space. This range of $M_{\rm hm}$ covers the most commonly considered WDM models, including the 3.5 keV model. The lower limit M$_{\rm hm} = 10^6$ M$_{\odot}$ is strictly speaking warmer than CDM, but in practice indistinguishable from it within the mass range probed by the sensitivity of the data (see Section~\ref{sec:dm_massfnct}). 

The posterior probability density distribution is obtained assuming the different lens systems in the sample to be statistically independent from each other. 

\section{Lens modelling results}
\label{sec:results}

A complete description of the analysis and the results of the smooth modelling is provided in \citet{Ritondale18}, along with a detailed comparison with the smooth models derived by \citet{Shu16a}.
In this section, we present the results of our search for low-mass haloes.

\subsection{Substructure search}
\label{sec:clumpyres}

Out of the 17 gravitational lenses in the sample, we find that 14 do not fulfil one or more of the detection criteria defined in Section~\ref{sec:criteria} and, therefore, all the pixels in these systems will contribute to the statistical analysis as non-detections. In particular, in all of these 14 cases, the smooth model is always preferred by the Bayesian evidence, independently of the choice of prior. Moreover, no significant convergence corrections are identified. For the remaining three systems, namely \rm{SDSS~J0742+3341}, {\rm{SDSS~J0755+3445} and \rm{SDSS~J1110+3649}, we find that the Bayesian Evidence persistently prefers a model that includes the presence of one or more subhaloes, however, the potential corrections give a meaningful perturbation only in the case of \rm{SDSS~J1110+3649}. Below, we discuss these systems individually in more detail.

\subsubsection{\rm{SDSS~J0742+3341}}

The parametric analysis of Section \ref{sec:clumpy}, where subhaloes are described as analytical NFW mass components, shows a persistent preference for a model that includes a substructure with a mass of $M_{\rm vir}^{\rm NFW} = (3.8 \pm 0.8)\times 10^{10} M_{\odot}$ located at $(dx,dy)= (1.14\pm0.04, -0.80\pm0.03)$ arcsec, relative to the main lens centre.
However, the detection is only at the 6-$\sigma$ level and therefore below our 10-$\sigma$ threshold (see Section~\ref{sec:tests}). Moreover, no significant and localized convergence correction is identified by the pixellated analysis of Section \ref{sec:potcorr} at the same location, as shown in Fig.~\ref{fig:pot_panel}. We, therefore, conclude that there is no evidence for a significant detection of a mass perturbation in this system and register it as another non-detection in the sample.

\subsubsection{\rm{SDSS~J0755+3445}}
\label{sec:j0755}

This system is a clear example of how a purely analytical analysis of mass substructure can lead to false detections due to a mis-modelling of the main lens macro model. We refer to Ritondale et al. (in prep) for an in-depth discussion of this system and related issues with the lens macro-model. Here, we provide a short summary of the results. The analytical clumpy analysis shows a consistent preference at the 12$\sigma$ level (i.e. $\Delta\log \mathcal{E} = 72$) for a model using a NFW halo with a mass of $M_{\rm vir}^{\rm NFW} = (4.8\pm0.4)\times 10^{10}$ M$_{\odot}$ located at $(dx,dy)= (-1.76\pm0.02, 1.12\pm0.02)$ arcsec, relative to the main lens centre. However, no strong positive and localized convergence correction is identified. Low-level diffuse corrections can be seen instead (see Fig.~\ref{fig:pot_panel}), which allow for a better focusing of the background source and reduce the positive and negative beating otherwise seen in the low-surface brightness tail of the smooth source. This indicates that the true mass distribution of this system is probably not well described by a single power-law model and that the compactness of the gravitational imaging source is probably due to the extended convergence corrections shown in Fig.~\ref{fig:pot_panel}. 

Moreover, we find that the Bayesian evidence further increases when adding a second analytic subhalo. The source tail is also further decreased, although we do not see evidence for this second halo in the gravitational imaging as a localized positive correction either. We have also modelled the data using a double power law model and including the contribution to the lensing potential of a galaxy observed in the south-west direction and $\sim$5 arcsec away from the main lens. However, in none of these cases could we reconstruct a compact source and remove the low-level diffuse potential corrections. We conclude therefore that complex mass components that remain un-captured by the macro-model can mimic the effect of subhaloes and lead to an overestimation of the latter. 

This result is in qualitative agreement with what was found by \citet{Hsueh16,Hsueh17}, who have shown that observed flux-ratio anomalies in multiply imaged quasars can sometimes be reproduced by the lensing effect of an un-modelled edge-on disk rather than requiring dark matter (sub)haloes. Similarly, from an analysis of hydrodynamical simulations, \citet{Hsueh18} have shown that the presence of baryonic structures and disks is responsible for an increase of flux-ratio anomalies by a factor from 8 to 20 per cent. Moreover, they found that baryonic structures can cause astrometric anomalies in 13 per cent of the studied mocks. \citet{Gilman17a} have come to the same conclusion by analysing mock data based on {\it HST} observations of low-redshift galaxies. 

This demonstrates that the gravitational imaging technique is important to distinguish between genuine detections and false detections due to an un-captured underlying complexity of the lensing mass distribution.

\begin{figure*}
\begin{center}
\subfloat[]{\includegraphics[width= 8.57 cm, keepaspectratio, viewport=39 20 585 426, clip]{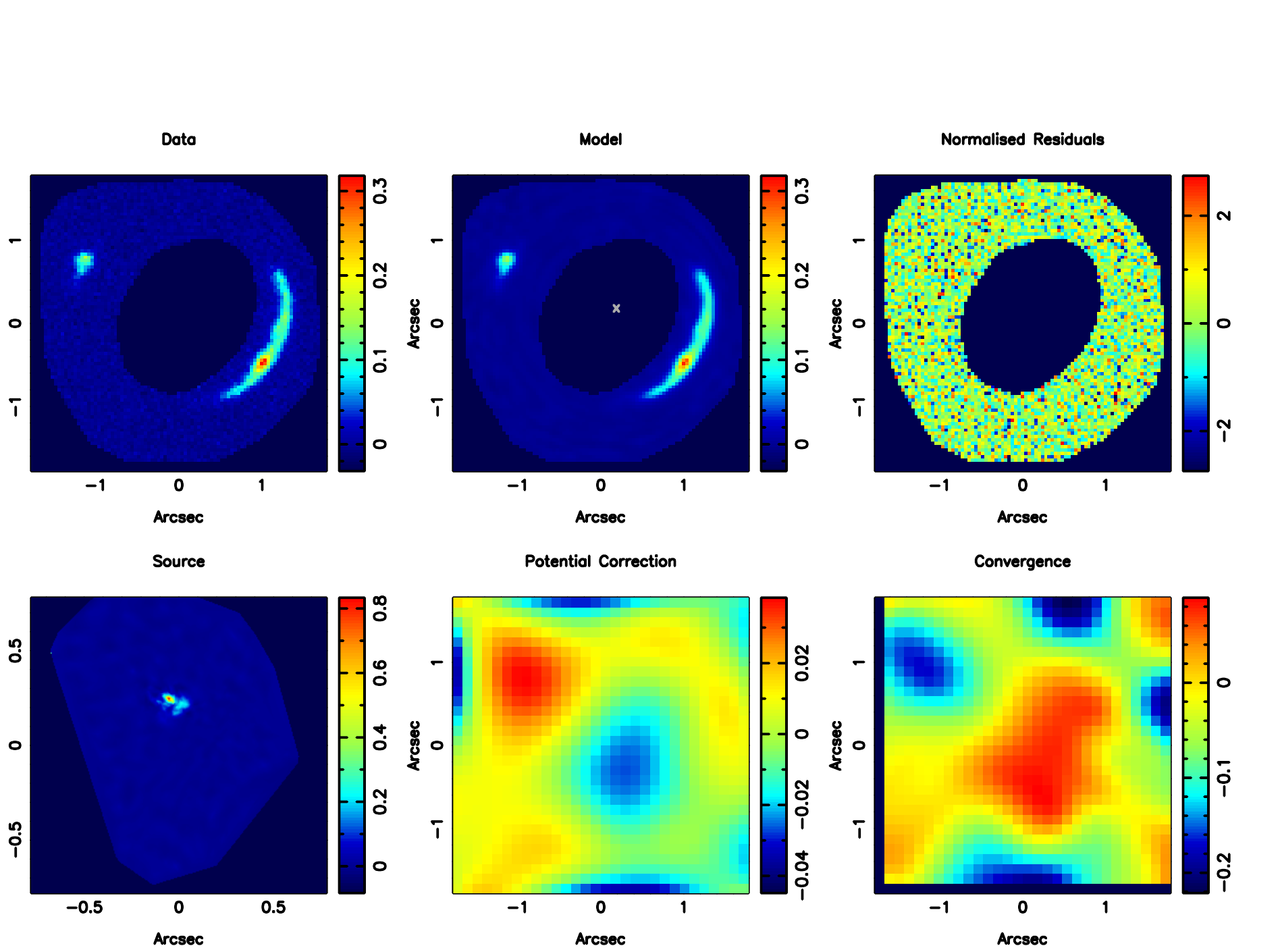}}\qquad
\subfloat[]{\includegraphics[width= 8.57 cm, keepaspectratio, viewport=39 50 565 426, clip]{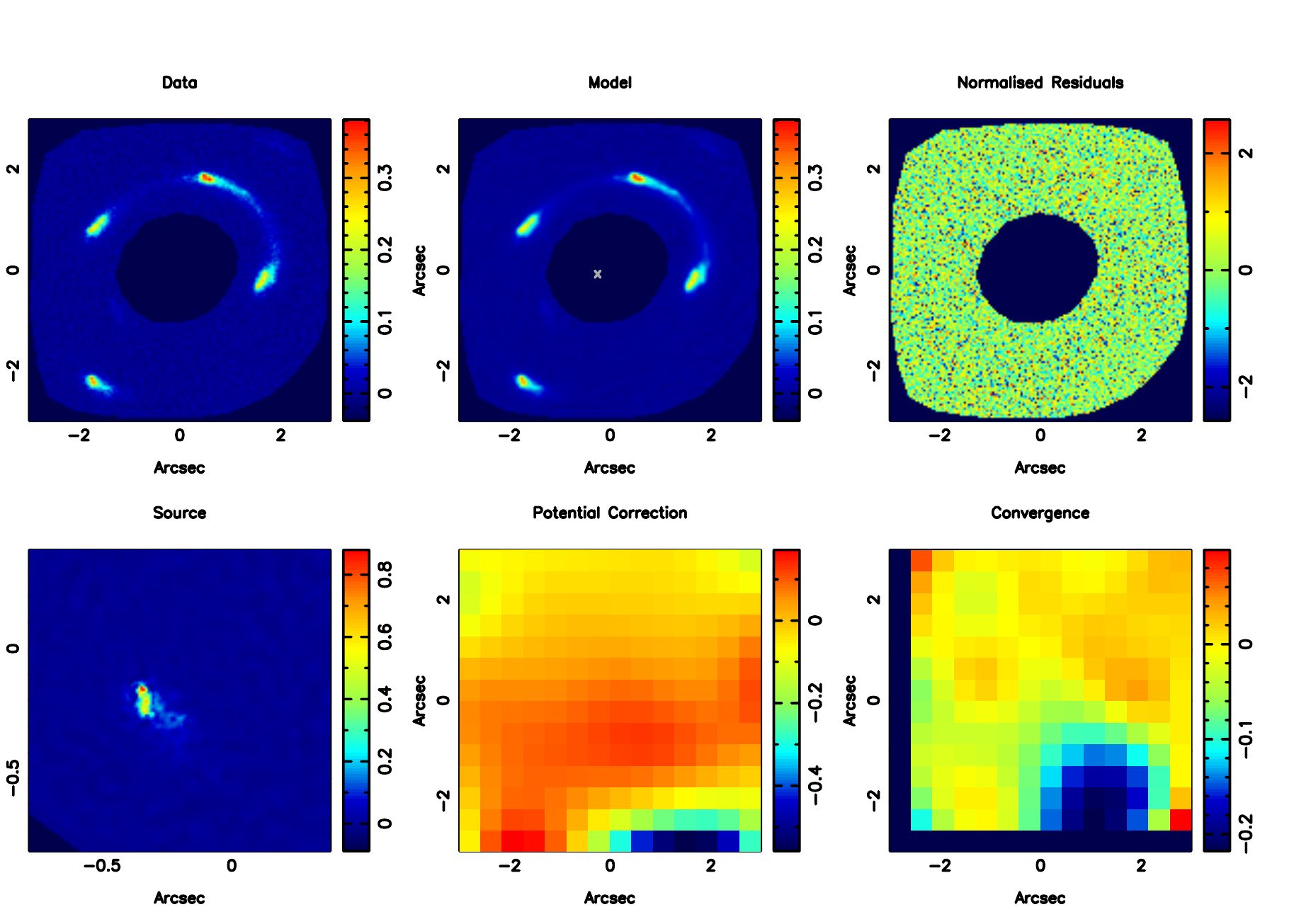}}\qquad
\subfloat[]{\includegraphics[width= 8.57 cm, keepaspectratio, viewport=39 50 565 426, clip]{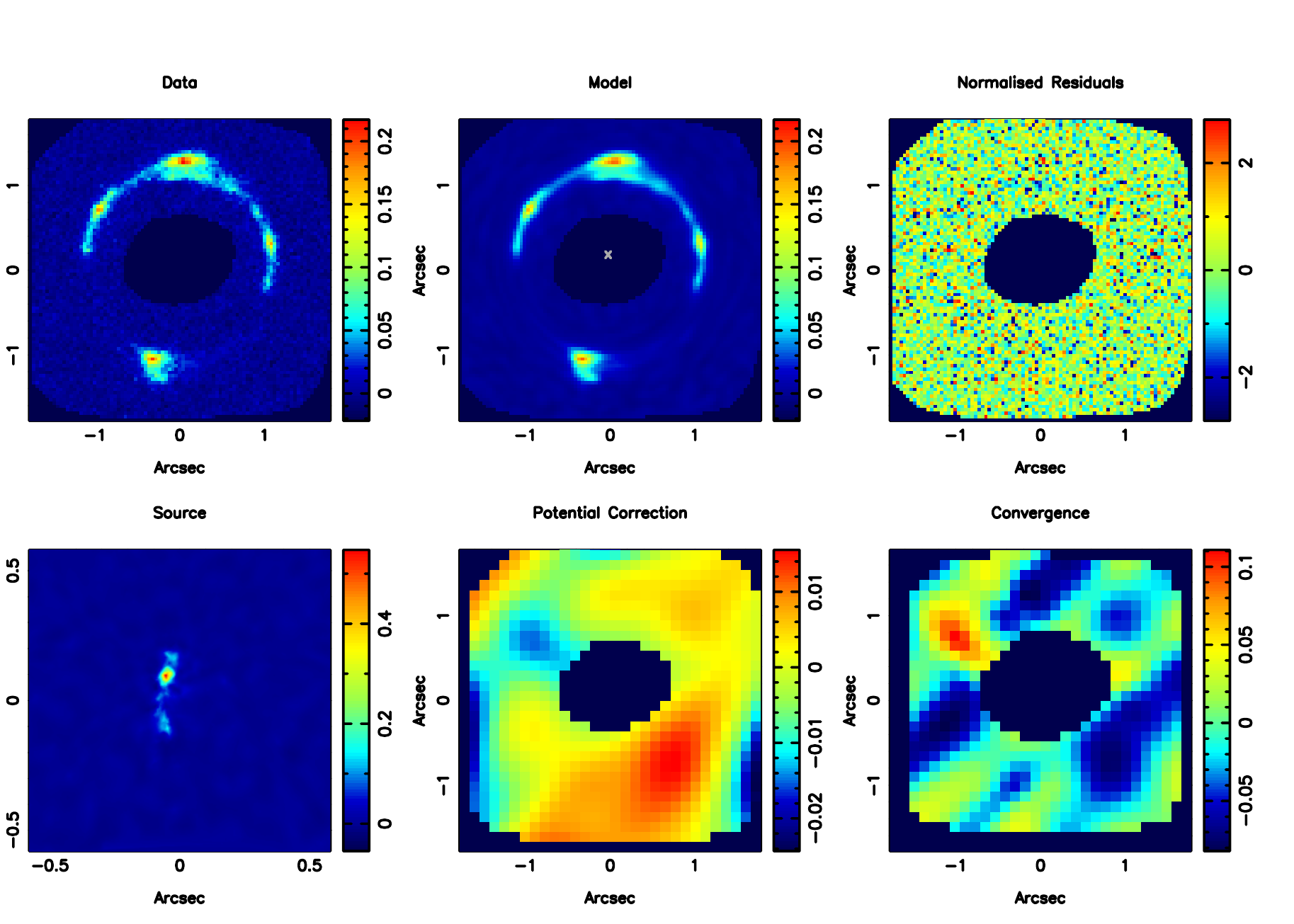}}\\
\caption{Results of the gravitational imaging analysis for the lens systems {\rm{SDSS~J0742+3341}} (Panel a), {\rm{SDSS~J0755+3445}} (Panel b) and {\rm{SDSS~J1110+3649}} (Panel c). For each lens, the top-row shows the data (left), the model (middle), the normalised residuals (right). The bottom-row shows, the reconstructed source (left), the pixellated potential corrections (middle) and the corresponding convergence corrections (right). In {\rm{SDSS~J0742+3341}} the convergence corrections are in correspondence of the lens galaxy, but their low-level and wide extension suggest the absence of small size haloes. {\rm{SDSS~J0755+3445}} also shows low-level and diffused corrections to the potential; this is a symptom of a complicated mass distribution rather than of the presence of a subhalo. In {\rm{SDSS~J1110+3649}} we see the presence of a positive and localized potential correction, possibly indicating the tentative detection of a dark matter halo of mass $M_{\rm 2D}^{\kappa}(<r_{\rm s}) = 2.5 \times 10^9 {\rm M}_\odot$.}
\label{fig:pot_panel}
\end{center}     
\end{figure*}

\subsubsection{\rm{SDSS~J1110+3649}}
The pixellated analysis for {\rm{SDSS~J1110+3649}} shows the presence of a positive localized convergence correction at about $(dx,dy)= ( -0.851, 0.628)$ relative to the main lens (Fig.~\ref{fig:pot_panel}). Moreover, the parametrized analysis of Section \ref{sec:clumpy} gives a preference at the 4-$\sigma$ level for a model with a subhalo with mass $M_{\rm{vir}}^{\rm NFW} = (5.4\pm1.5)\times10^9 M_\odot$. This subhalo is located at $(dx,dy)= ( -0.945 \pm 0.084, 0.536 \pm 0.021)$ relative to the main lens centre, consistent within 1$\sigma$ with the position of the convergence correction. From the pixellated convergence corrections, we derived a model-independent projected mass for the substructure of $M_{\rm 2D}^{\kappa}(<r_{\rm s}) = \sum_{\substack{pix}}~2~\pi~\Sigma_c~\kappa_i~r^2_i =  2.5 \times 10^9 {\rm M}_\odot$. This is consistent within 2$\sigma$ with the parametric projected mass $M_{\rm 2D}^{\rm NFW}(<r_{\rm s}) = (1.7\pm 0.7) \times 10^9 {\rm M}_{\odot}$.

However, given the low statistical significance and the results presented in Section~\ref{sec:tests}, we conclude that multi-band data or deeper observations are required to understand the nature of this system and, at present, also in this case we do not count this as a detection. 

\section{Inference on the dark matter parameters}
\label{sec:inferencedm}
In this section, we combine the lens modelling results presented in Section \ref{sec:results} with the statistical formalism introduced in Section \ref{sec:stats} to derive statistical constraints on the free streaming properties of dark matter.
First, we compare with the expected value of detectable line-of-sight haloes from the CDM paradigm \citep[i.e. $\alpha = 1.9$, $M_{\rm hm}=0$,][]{Springel2008} and then with resonantly produced sterile neutrino models \citep{Shi99} including the contribution of both subhaloes and line-of-sight haloes.

\subsection{Sensitivity function}
\label{sec:mu}

We firstly compute the sensitivity function for the BELLS GALLERY sample, as described in Section~\ref{sec:sensitivity}. As discussed in the same section, the high level of structure of the sources could in principle provide a high sensitivity to low-mass haloes at fixed signal-to-noise. However, this was found not to be the case: the BELLS GALLERY lenses not only have a mean sensitivity that is lower than the SLACS lenses (see Fig.~\ref{fig:sens}), but also, as the background sources are very compact, have a smaller fraction of image plane pixels with a high sensitivity than SLACS. For this reason, this sample of lenses turned out to be less constraining than the SLACS lenses in terms of probing the halo and subhalo mass function at an interesting mass regime. Higher signal-to-noise ratio observations are required to improve the sensitivity of this sample.

\subsection{A potential discrepancy with CDM}
\label{sec:CDMdiscr}
Assuming our reference detection threshold of 10$\sigma$ and the relative sensitivity function, we compute the number of detectable line-of-sight CDM haloes to be $\mu_{l} = 1.17\pm1.08$, for the complete sample of 17 systems, in agreement with the zero detections registered for this sample. This is computed with equation~(\ref{eq:mu}) using the the lowest detectable (at the 10-$\sigma$ level) mass in each pixel as the lower integration limit and summing over all 17 lenses. This result is consistent with the fact that this sample has relatively low sensitivity (i.e. large value for the lowest detectable mass) and thus the number of line-of-sight haloes per arcsec or per pixel is relatively small.

Although the tests presented in Section~\ref{sec:tests} have shown that a detection threshold cut at the 5-$\sigma$ level is not reliable, it is interesting to test what happens if the sensitivity improved to the level implied by this less conservative threshold. As can be seen in the bottom panel of Fig.~\ref{fig:sens}, the sensitivity at the 5-$\sigma$ cut of the BELLS GALLERY sample is closer to the 10-$\sigma$ level one of the SLACS lenses, with a tail at lower masses. As a consequence of the improved sensitivity, the number of detectable CDM line-of-sight haloes significantly rises to $\mu_l = 9.0\pm3.0$, while we find that also at the 5-$\sigma$ level, the number of detections in the sample is still zero. The unreliability of this sensitivity cut doesn't allow us to draw robust conclusions, but the probability of registering zero detections in the CDM framework would be P$^{{\rm5}\sigma}_{{\rm CDM}}(n_{\rm det}=0)$=0.0001. Interestingly, \cite{V18} have found that the expected number of CDM line-of-sight haloes at the 10-$\sigma$ level for the SLACS lenses is $\mu_l = 0.8\pm0.9$ \citep[in agreement with the single detection reported by][]{V14a}, reflecting the lower size of the cosmological volume probed by this sample. These results indicate that deeper exposures or multi-band data, that provide improved sensitivity whilst keeping the robust 10-$\sigma$ threshold, for the BELLS GALLERY sample hold significant promise to find a possible strong tension between the CDM model and the sample of lenses considered in this paper.

\begin{figure}
\begin{center}
\includegraphics[width= 9.5cm]{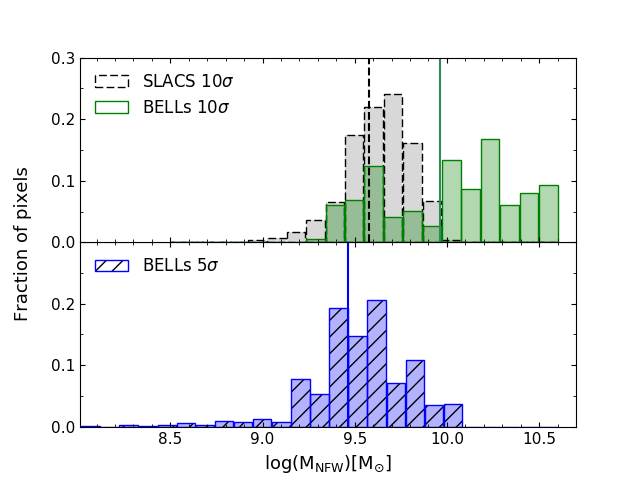}
\caption{The upper panel shows the fraction of pixels with a 10-$\sigma$ substructure mass detection threshold (as defined in Section \ref{sec:sensitivity}) for the BELLS GALLERY sample analysed in this paper (green histogram) and for the sub-sample of SLACS lenses analysed by \citet[][grey histogram]{V14a}. The lower panel shows the same for the BELLS GALLERY sample but with a 5-$\sigma$ substructure mass detection threshold. The vertical lines indicate the mean sensitivity values for each sample.}
\label{fig:sens}
\end{center}     
\end{figure}

\begin{figure}
\begin{center}
\includegraphics[height=5cm,keepaspectratio,viewport=-80 0 1080 736, clip]{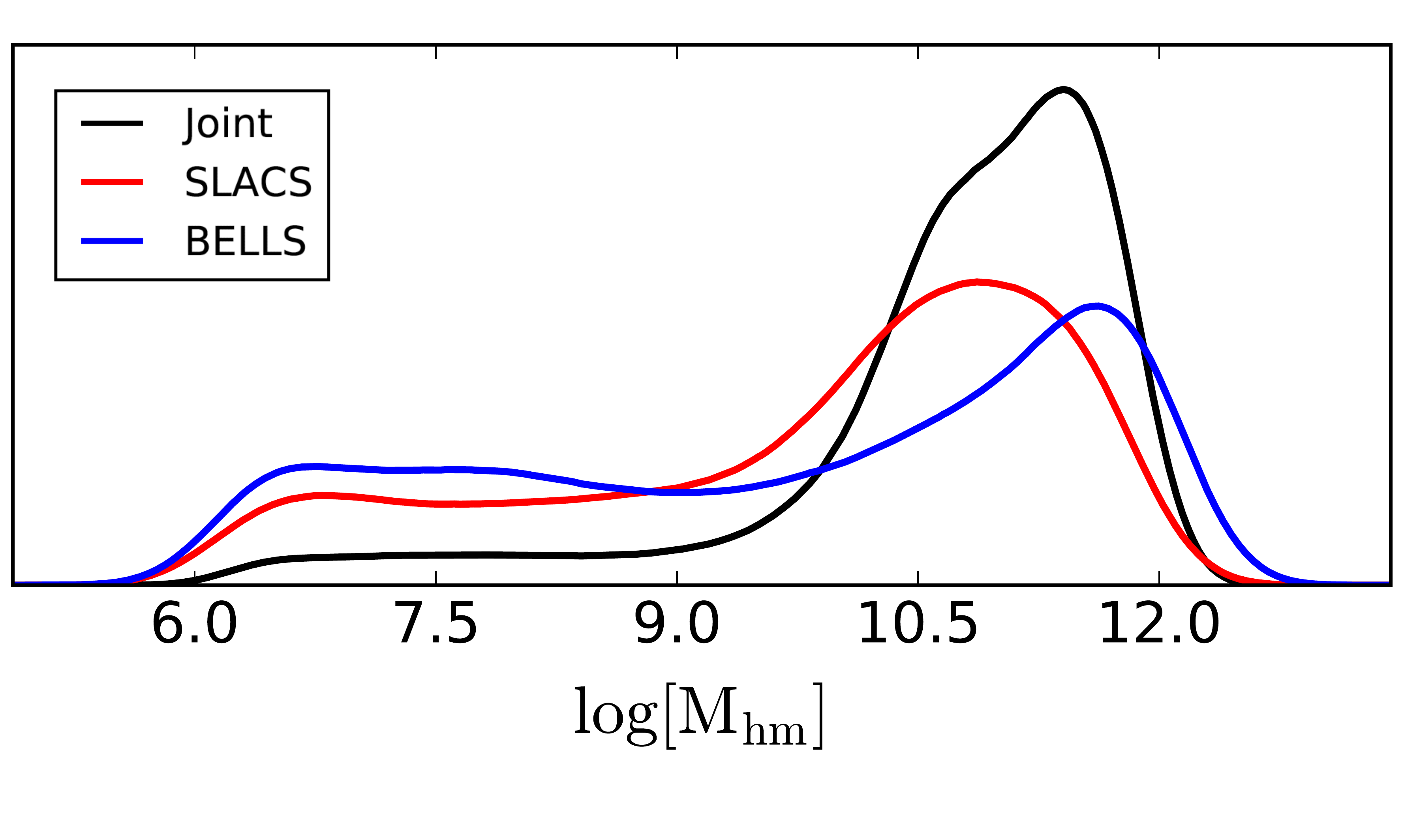}
\caption{The posterior probability density distribution for the half mode mass $M_{\rm hm}$ for the joint and individual samples.}
\label{fig:cornerplot}
\end{center} 
\end{figure}

\subsection{Dark matter mass function}
\label{sec:dm_massfnct}

In the previous section, we only looked at the total expected number of line-of sight haloes, here we use the full results of the Bayesian analysis (with the 10-$\sigma$ level cut) to characterize the dark matter model based on the total number of detections and non-detections  and using the priors described in Sections~\ref{sec:prior}.
We summarize our constraints on the subhalo and line-of-sight halo mass function parameters in Table~\ref{tbl:constr}. Specifically, we report the mean, and the upper and lower limits at the 68 and 95 per cent confidence level for the line-of-sight and substructure mass function slopes $\alpha$ and $\beta$, the 68 and 95 per cent upper limit on the dark matter mass fraction in subhaloes $f_{\rm sub}$, and the 68 per cent and 95 per cent level upper and lower limits for the half-mode mass $M_{\rm{hm}}$. The posterior probability distributions for these last two parameters obtained with the BELLS GALLERY sample are presented in Fig.~\ref{fig:cornerplot}. We have constrained the half-mode mass to be $\log M_{\rm{hm}}<12.60$ at the 2-$\sigma$ level. As expected from the calculations in Section~\ref{sec:mu}, our results are in agreement with the CDM paradigm, but do not allow us to rule out alternative warmer dark matter models.

Recently, \cite{V18} have performed a similar analysis with a sample of 11 gravitational lens systems from the SLACS survey by combining the single detection of \cite{V10} with the non-detections reported by \cite{V14a}. The substructure mass fraction derived here is smaller than the value reported by \cite{V18}, which is contrary to what one would expect given the relative redshift of the two samples of lenses (high redshift for the BELLS GALLERY and low redshift for the SLACS sample), however this is just a reflection of the poor data sensitivity, as shown in Fig.~\ref{fig:sens}, and the small number statistics combined with the null detection. Also, it should be noted that the definition of $f_{\rm sub}$ adopted here (total mass fraction in subhaloes) is different than the one adopted by \citet[][dark matter mass fraction in subhaloes]{V14a,V18}.

In Fig.~\ref{fig:cornerplot}, we plot the joint posterior probability distribution for $M_{\rm hm}$ derived by combing a posteriori the analysis of the two sample of lenses. It has to be noted that we do not provide a joint inference on $f_{\rm sub}$ as its definition is different for the two samples and it is expected to change with the mean lens redshift of the sample \citep{Xu15}. In Table~\ref{tbl:constr} we also show the 95 and 68 per cent upper and lower limits on the half-mode mass derived from the joint analysis of the SLACS and BELLS GALLERY samples. It is evident that the constraints at the lower 95 per cent confidence limits are driven by the SLACS sample, while the upper limits are driven by the BELLS GALLERY sample and are now consistent with warmer models, that is the joint 95 per cent upper limit has shifted towards larger values from what was derived using the SLACS sample only. This can be explained as follows: when combining the two samples, the number of detections is the same, that is one, while the number of non-detections significantly increases with the number of pixels in each lens system included in the analysis. The substructure detection in the SLACS sample is also responsible for a significant change in the inference on the half-mode mass $M_{\rm hm}$. In fact, the lower limit on $M_{\rm hm}$ raises by 3 and 1 dex at the 68 and 95 per cent confidence level, respectively. This is due to the fact that the single, rather massive detection from the SLACS sample requires a cooler dark matter model, and therefore smaller values of $M_{\rm hm}$, as clearly visible in the derived posterior probability in Fig.~\ref{fig:cornerplot}.

In Fig.~\ref{fig:massfun}, we compare the differential line-of-sight halo mass function derived in this paper with the one predicted by the CDM model (black solid line) and a sterile neutrino model consistent with the 3.5 keV emission line (red solid line). The latter falls within our lower and upper 95 per cent confidence limits, respectively plotted as the green and yellow solid lines. Our lower limit mass function is consistent at the 2-$\sigma$ level with the CDM prediction within the mass range probed by the data. The inability to disentangle CDM and warmer models, is due to the relatively low sensitivity of the data to low-mass haloes, represented by the grey shaded region. In practice, this data can only probe the higher-mass end of the halo and subhalo mass functions, where different dark matter models do not significantly differ from one another \citep{Despali18}. As discussed by \cite{V18}, the same sample of lenses with a sensitivity improved by one or two orders of magnitude would result in a shift of the posterior distribution of the half-mode mass towards larger values and create a tension with CDM at the 2-$\sigma$ level. This clearly indicates the importance of obtaining higher-quality data for the joint sample.

In Fig.~\ref{fig:mvsL6andexclreg}, we show how our results compare with sterile neutrino dark matter models. Sterile neutrinos are a two-parameter dark matter model whose \emph{coolness} is determined by a combination of the level of lepton asymmetry $L_{\rm 6}$ in the early Universe and the mass of the sterile neutrino $m_{\rm s}$ \citep{Shaposhnikov08,Lovell17a}. This is evident from Fig.~\ref{fig:mvsL6andexclreg}, where $M_{\rm{hm}}$ oscillates with $L_{\rm 6}$ for each value of $m_{\rm s}$. On the left panel of Fig.~\ref{fig:mvsL6andexclreg} we plot the half-mode mass $M_{\rm{hm}}$ against the lepton asymmetry L$_{\rm6}$ for different values of the sterile neutrino particle mass. On the right panel instead, we compare our results with those derived from the observed satellites in the Milky Way \citep{Lovell16}, X-ray decay searches from M31 \citep{Watson12,Horiuchi14} and Lyman-$\alpha$ forest constraints \citep[see][for a detailed description]{V18}. We notice that our joint lower limit constraints are not visible because they are beyond the plotting range. The upper 95 per cent confidence limit rules out sterile neutrino masses $m_{\rm s} < 0.8$ keV at any value of the lepton asymmetry $L_{\rm 6}$. As for the SLACS-only results, our exclusion regions are significantly smaller than those derived by other astrophysical probes. For the SLACS lenses, this is mainly due to the low redshift of the lenses and the sources, which results in a small contribution from the line-of-sight. For the BELLS GALLERY sample, this is instead related to the lower sensitivity of the data. Indeed, as discussed in Section \ref{sec:mu}, the same sample of lenses but with higher data quality than currently available would have led to a significantly larger number of expected line-of-sight haloes.
Finally, it should be noted that, although our results are currently weaker, they are more robust than those from the Milky-Way satellite counts and the Lyman-$\alpha$ forest, as they are less affected by feedback processes and do not depend on the unknown thermal history of the intergalactic medium, and our limits are therefore less model dependent. 

\begin{figure}
\begin{center}
\includegraphics[height=7cm,keepaspectratio]{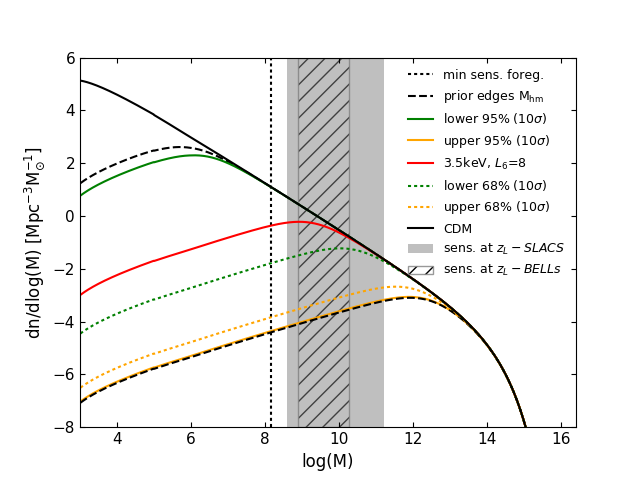}
\caption{Line-of-sight mass functions derived from the joint SLACS+BELLS GALLERY dataset. The black line corresponds to the $\Lambda$CDM framework, the red to the sterile neutrino dark matter model compatible with the detection of the 3.5 keV line, and the yellow and green respectively to the upper and lower limits at the 95 (solid line) and 68 (dashed line) per cent confidence levels found in this paper, assuming a 10-$\sigma$ detection threshold. The black dashed lines correspond to the prior edges in $M_{\rm hm}$. The striped and shaded grey regions correspond to the sensitivity of the BELLS GALLERY and the SLACS samples, and the dotted line shows the lowest detectable line-of-sight halo mass with the joint sample.}
\label{fig:massfun}
\end{center} 
\end{figure}

\begin{figure*}
\begin{center}
\subfloat[]{\includegraphics[height=6.4cm,keepaspectratio]{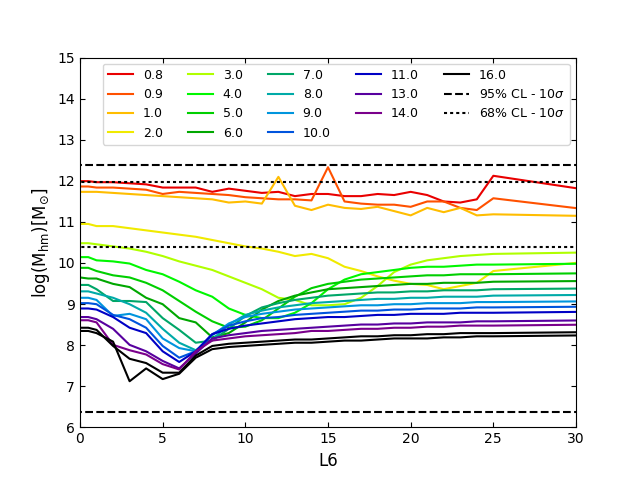}}\qquad
\subfloat[]{\includegraphics[height=6.4cm,keepaspectratio]{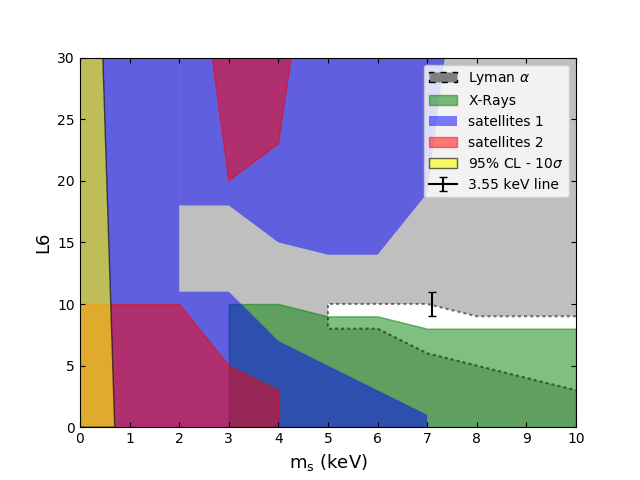}}
\caption{Left: Half-mode mass versus lepton asymmetry $L_{\rm 6}$ for different values of the sterile neutrino mass with upper and lower limits at the 95 and 68 per cent confidence level on the turnover mass (dashed and dotted black lines). Right: 95 per cent exclusion region in the $L_{\rm 6}$ versus sterile neutrino mass $m_s$ plane. The grey region has been excluded by the Lyman-$\alpha$ forest, the green region is excluded by the missed detection of X-ray decay in Andromeda, and the blue and red regions are excluded by satellite counts in the Milky Way with two different feedback models. The yellow region is excluded at the 95 per cent confidence level by the detections and non-detections in the joint SLACS+BELLS GALLERY sample. Only our upper limit is visible, since the lower limit we set lies outside the mass range of this plot, being much higher than the masses constrained by other methods. The black error bar corresponds to the dark matter model explaining the 3.5 keV line.}
\label{fig:mvsL6andexclreg}
\end{center} 
\end{figure*}

\begin{table}
\caption{Inference on the dark matter parameters with the BELLS sample and the joint BELLS and SLACS samples. We report the mean and the lower and upper limit at the 68 and 95 per cent confidence level for the two mass function slopes, while we only report the upper and lower limits for the half mode mass $M_{\rm{hm}}$ and the upper limits on the dark matter fraction in substructures at the 68 and 95 per cent level.}
\label{tbl:constr}

\begin{tabular}{ccccc}
\hline
Run & Parameter  & mean & $\sigma_{\rm{68}}$ & $\sigma_{\rm{95}}$ \\ 
\hline
BELLS & $\alpha$ & 1.90 &  $-$0.19 | $+$0.19   &   $-$0.33 | $+$0.37 \\
 &  &  &  &  \\
 &  $\beta$ & $-$1.30 & $-$0.1 | $+$0.09      &   $-$0.16 | $+$0.16\\
  &  &  &  &  \\
 &  $f_{\rm{sub}}$ &   & <0.01       &  <0.07 \\
  &  &  &  &  \\
 &  log$M_{\rm{hm}}[M_{\odot}]$ &   & 6.52 | 12.12  &  5.77 | 12.60\\
\\
Joint &  log$M_{\rm{hm}}[$M$_{\odot}]$ &   & 10.38 | 11.85  &  7.27 | 12.26\\ 
\hline
\end{tabular}
\end{table}

\section{Summary \& conclusions}
\label{sec:conclusions}

We have analysed a sample of 17 gravitational lens systems from the BELLS GALLERY survey with the aim of detecting low-mass dark matter haloes within the lensing galaxies and along their lines of sight. First, we have modelled each system in the sample with a smooth power-law elliptical mass model assuming the presence of no haloes and studied the intrinsic properties of the background sources \citep{Ritondale18}. In this paper, we have focused on the detection of low-mass haloes and its implication for the dark matter properties, in particular those of sterile neutrinos.

Our main results can be summarised as follows. For the entire sample of lenses, we report no significant detection of subhaloes and line-of-sight haloes. In particular, 14 systems show statistical preference for a model that does not include the presence of any halo; one system, {\rm{SDSS~J0742+3341}}, shows a small preference for a model with a subhalo, however this was not confirmed by our gravitational imaging analysis. The system {\rm{SDSS~J1110+3649}} also shows a preference for a small-mass subhalo with a corresponding significant pixellated convergence correction. However, its statistical significance is still below our 10-$\sigma$ detection threshold and we, therefore, conclude that more data is required to draw final conclusions on this system. 

\citet{Hsueh16,Hsueh17} have shown that un-modelled edge on-disks and other baryoninc structures in early-type galaxies can cause flux-ratio and astrometric anomalies in multiply imaged quasars, similarly to dark matter (sub)haloes. A similar conclusion was reached using realistic lens galaxies taken from numerical simulations and mock data based on {\it HST} observations of low-redshift galaxies \citep{Hsueh18, Gilman17a}. On the same note, our analysis of the lens system {\rm{SDSS~J0755+3445}} has shown how structure not readily visible in the imaging data might affect the lensing and therefore how complex mass distributions can potentially lead to the false detection of mass substructure. This result is particularly important to show how a pixellated gravitational imaging analysis can be used to distinguish between the two scenarios.

Assuming a sensitivity function for the detection of subhaloes based on a 10-$\sigma$ cut and applying the mass-redshift relation by \citet{Despali18}, we have derived the total expected number of CDM line-of-sight haloes for our entire sample to be $\mu_{l} = 1.17\pm1.08$, in agreement with our null detection. Our results are therefore consistent with the $\Lambda$CDM model, under our most conservative assumption on the subhalo detectability. 
Interestingly, if we were to relax our assumptions and adapt a 5-$\sigma$ cut, the number of detectable line-of-sight haloes raises significantly to $\mu_{l} = 9.04\pm3.01$, that would potentially be in strong tension with our results that point to zero detections also at this sensitivity cut. However, we have extensively tested our detections and non-detections at the 5-$\sigma$ confidence level and we have found a high percentage of false positive and false negatives. We conclude therefore that the currently available data for this sample does not allow us to draw robust conclusions at the 5-$\sigma$ level. Therefore, deeper exposure (we note that the sensitivity improves non linearly with the data quality) or multi-band data are required to improve the sensitivity whilst keeping the robust 10-$\sigma$ threshold and consequently to find a potential strong discrepancy with the standard cosmological model.

We have used the BELLS GALLERY lenses to infer the dark matter mass function and constrained the half-mode mass to be $\log M_{\rm{hm}}<12.60$ at the 2-$\sigma$ level. If we combine our results with those derived by \cite{V18} from a subsample of the SLACS lenses, our constraints drop to $\log M_{\rm hm}<12.26$ at the same confidence level. An interesting result is the significant change in the inference on the dark matter model parameters when even just one detection is included. In fact, we register a shift of 3 dex on the 68 per cent lower limit for $M_{\rm hm}$ after combining the two samples. More sensitive data is necessary to significantly improve the constraints at the 95 per cent confidence level.

Assuming that the dark matter is composed by resonantly produced sterile neutrinos, we have then derived a 95 per cent confidence level exclusion region for the sterile neutrino mass and the lepton asymmetry in the early Universe, which is significantly smaller than the constraints obtained with other astrophysical probes, such as the number of Milky Way satellites and the Lyman-$\alpha$ forest. Specifically, our current results are consistent with the CDM paradigm, but do not allow us to rule out alternative warmer dark models. This is due to the limited sensitivity of the current data, which only allows us to probe the high-mass end of the dark matter mass function, where different dark matter models predict a similar number density of subhaloes and line-of-sight haloes.

In the future, observations of strong gravitational lens systems with a redshift distribution similar to the BELLS GALLERY sample considered here, but with a higher data quality (i.e. higher signal-to-noise ratio) will allow us to set tighter and robust constraints on the nature of dark matter.
 



\section*{Acknowledgements}
The authors are grateful to Mark Lovell and Simon D. M. White for useful discussions. This work is based on observations made with the NASA/ESA Hubble Space Telescope ({\it HST}). S.~V. has received funding from the European Research Council (ERC) under the European Union's Horizon 2020 research and innovation programme (grant agreement No 758853). LVEK is partly supported through an NWO-VICI grant (project number 639.043.308).

\bsp	
\label{lastpage}

\bibliographystyle{mnras}
\bibliography{ms}

\end{document}